\documentclass{XrU2005}

\title{Temperature and Dark Matter Profiles of Galaxy Groups}
\author[1]{F. Gastaldello}
\author[1]{D. Buote}
\author[1]{P. Humphrey}
\author[1]{L. Zappacosta}
\author[1]{J. Bullock}
\author[2,3]{F. Brighenti}
\author[2]{W. Mathews}
\affil[1]{Department of Physics and Astronomy, University of California Irvine}
\affil[2]{UCO/Lick Observatory, Board of Studies in Astronomy and Astrophysics, University of California Santa Cruz}
\affil[3]{Dipartimento di Astronomia, Universit\`a di Bologna}

\newcommand\cvir{{\hbox{$c_{\rm vir}$}}}
\newcommand\chandra{{\sl Chandra}}
\newcommand\xmm{{\sl XMM}}
\newcommand\mvir{{\hbox{$M_{\rm vir}$}}}
\newcommand\rvir{{\hbox{$r_{\rm vir}$}}}
\newcommand\msun{\hbox{{$M_{\odot}$}}}
\usepackage{graphicx}

\begin{document}

\keywords{ X-rays: galaxies: clusters;  galaxies: halos; dark matter}

\maketitle

\begin{abstract}
\vspace{-0.5cm}
The \chandra\ and \xmm\ data for a sample of 19 relaxed groups/poor clusters,
covering the temperature range 1-3 keV and selected to have the best available
data, reveal a remarkable similarity in their temperature profile: cool cores
outside of which the temperatures reach a peak for radii less than 0.1 \rvir\ 
and then decline.\\
We fitted the derived mass profiles using an NFW model, which provides 
a good fit to the data when accounting for the central galaxy in the inner region.
The concentration parameters and virial masses are in the range \cvir = 5-22 and \mvir = 
$2 \times 10^{13} - 4 \times 10^{14}$ \msun, in general agreement with the concentrations found in numerical simulations.  

\end{abstract}

\vspace{-0.5cm}
\section{Introduction}
\vspace{-0.5cm}
The nature of Dark Matter (DM) in the universe is one of the fundamental 
problems in astrophysics and cosmology. Crucial is the comparison of
observation with N-body simulations in the currently favored $\Lambda$CDM
cosmology, which predicts a universal NFW profile for DM halos \citep{nfw}.
This prediction has been tested for the scale of hot, massive clusters 
\citep{point05,vikh05} but few constraints exist on the group scale, where
predictions for the statistical properties of DM halos are more accurate 
because a large number of objects can be simulated at once \citep[e.g.][]{bullock01}.
The distribution of concentration parameters ($c$) is expected to vary 
significantly as a function of the cosmological parameters \citep[e.g.][]{kuhlen05}.
A measure of the mean and scatter of $c$ at the group scale is a crucial test
of the $\Lambda$CDM model.

The inner halo profiles also reveal vital information about the interplay between the DM 
and baryons during group formation. It is expected that in the central region 
of group halos the pure NFW profile will be modified by adiabatic compression
as a response to baryonic dissipation \citep[][]{blum86}. This raises the question
whether the DM profile at the center of groups should be represented by a pure
or modified (adiabatically compressed, AC) NFW profile. DM properties can strongly constrain the galaxy formation process.

\vspace{-0.5cm}
\section{Data analysis}
\vspace{-0.5cm}
\subsection{Sample selection}
\vspace{-0.5cm}
The sample has been selected to cover the range in the temperature from 1 to 3 keV, the 
temperatures expected for groups/poor clusters, with as much as possible regular X-ray 
morphology and 
with a dominant elliptical galaxy at the center (the only exception being RGH80, with two 
ellipticals of 
comparable size at the center), to ensure dynamical relaxation.
\vspace{-0.5cm}
\subsection{Background subtraction}
\vspace{-0.5cm}
A correct background subtraction is a critical element of our analysis. We describe here the 
procedure adopted for XMM. 
Soft proton flares cleaning have been performed on a region free of the luminous core of the 
objects and point sources and using both a hard and 0.5-10 keV light curve. 
The background has been then entirely modeled with a procedure similar to \citet{lumb02}. 
This method is particularly effective in studying groups because the source component (mainly 
characterized by the Fe-L blend) is clearly spectrally separated from all the other 
background components. The model has been applied to a simultaneous fit of source+background 
in the outer annuli to correctly determine the normalizations of the various background 
components.
\vspace{-0.5cm} 
\subsection{From temperature and density profiles to mass profiles}
\vspace{-0.5cm}
Spectra were extracted in concentric circular annuli located at the X-ray centroid and 
fitted with an APEC modified by Galactic absorption.  We projected parameterized 3D models 
for gas density and temperature to fit the results obtained from our spectral analysis 
(the projected gas mass density is derived from the APEC normalization). 
Using the best fit parameterized function we derive the total gravitating mass under the 
assumption of spherical symmetry and hydrostatic equilibrium. The virial quantities quoted in 
the poster are for an over-density of $\approx 101$ appropriate for the $\Lambda$CDM model.

\vspace{-0.5cm}
\section{Temperature profiles}
\vspace{-0.5cm}

\begin{figure}
\centering
\includegraphics[angle=-90,width=0.8\linewidth]{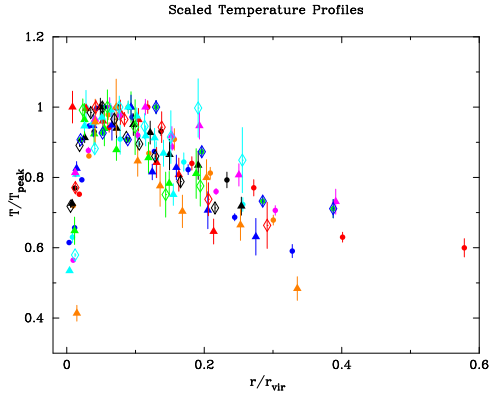}
\caption{\xmm\ temperature profiles for the sample normalized by the peak in the temperatures 
profile for each cluster and scaled by the virial radius obtained by the NFW fit.  
AWM 4 is shown with red triangles.\label{fig:1}}
\end{figure}

The temperature profiles for the systems, scaled by the virial radius obtained in the NFW fit, 
show a remarkable similarity. The shape resembles the one obtained for hotter, more massive 
clusters \citep[e.g.][]{vikh05} but with the peak of the temperature profile occurring at 
smaller radii (see Fig.\ref{fig:1}). 
The central cooling region is smaller and can be extremely affected by non-gravitational 
processes like radiative cooling and AGN heating \citep[as it is probably the case for AWM 7 
which has an interesting isothermal central profile,][]{sull05}.

\vspace{-0.9cm}
\section{A case study: NGC 1550}

\begin{figure}
\centering
\includegraphics[angle=-90,width=0.5\linewidth]{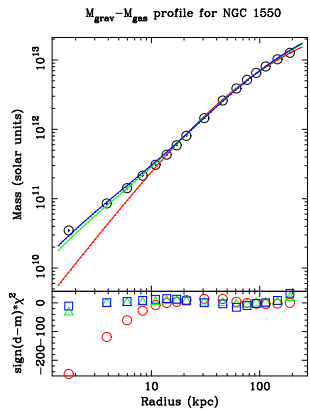}
\caption{DM+stellar mass fitted by a pure NFW (red), an NFW+Hernquist profile (green), an AC 
NFW model (blue) and residuals in the same colors.\label{fig:2}}
\vspace{-0.5cm}
\end{figure}

\vspace{-0.5cm}
As an example of the procedure adopted for each object in the sample we show here the case of 
NGC 1550. We analyzed the two ACIS-I and XMM observations and fitted jointly the temperature 
and density profiles. The derived gravitating mass profile has been fitted by an NFW profile. 
The DM+stars profile (obtained by subtracting the gas mass from the gravitating mass) 
has been fitted by a pure NFW, an NFW + stellar component for the central galaxy, with the latter modeled by an 
Hernquist profile \citep{hern90} or an AC model \citep{gnedin04}, as shown in Fig.\ref{fig:2}.

\vspace{-0.6cm}
\section{Variation of concentration with mass}
\vspace{-0.6cm}

\begin{figure}
\centering
\includegraphics[angle=-90,width=0.9\linewidth]{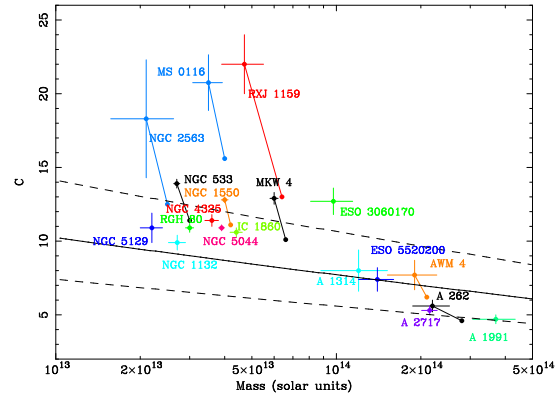}
\caption{Concentration parameters as a function of mass. The solid line shows the mean c(M) 
and outer lines the 1$\sigma$ scatter from numerical simulations \citep{bullock01}. The 
lines connect to the end point (shown without error bars) in the plot for objects requiring the 
introduction of a stellar component.\label{fig:3}}
\end{figure}

The resulting relation between $c$ and group mass is shown in Fig.3. An NFW profile is 
a good representation of the mass profiles observed; for 8 objects an excess compared to the 
NFW fit is present in the inner regions (as in NGC 1550): when fitted with an NFW+Hernquist or 
AC model, with the stellar mass free to vary, the returned stellar $M/L_B$ for the central galaxy are in the range 2-7. We can 
not discriminate between these two latter models. The inferred $M/L$ are considerably lower 
than the ones measured using stellar kinematics \citep[e.g.][]{gerhard01}.
The inclusion of the stellar component in these objects has the effect of lowering $c$ and 
increasing the mass.
The typical values and scatter of concentrations are in general agreement with the simulation 
results, as shown in Fig.\ref{fig:3}. Even after the inclusion of adiabatic contraction the 
results tend to be higher than predicted: a possible explanation is that we are looking 
at a biased population of groups \citep[relaxed, with some objects in our sample  
being classified as fossil groups, which should show an higher $c$ being formed at earlier 
epochs,][]{zentner05}.

\vspace{-0.6cm}
\section*{Acknowledgments}
\vspace{-0.5cm}
We thank O. Gnedin for kindly providing his AC code.

\vspace{-0.1cm}
{\footnotesize
\bibliographystyle{XrU2005}
{}
}
\end{document}